\documentclass{easychair}
\usepackage[utf8]{inputenc}
\usepackage{authblk}
\usepackage{cite}
\usepackage{authblk}
\usepackage{hyperref}
\usepackage{xcolor}
\usepackage{xspace}
\usepackage{mdframed}

\newcommand{\lazybvtoint}{\texttt{lazybv2int}\xspace}
\newcommand{\smtcomp}{SMT-COMP\xspace}

\newcommand{\qfbv}{QF\_BV\xspace}
\newcommand{\qfufnia}{QF\_UFNIA\xspace}
\newcommand{\msat}{MathSAT5\xspace}
\newcommand{\cvcfour}{CVC4\xspace}
\newcommand{\smtswitch}{smt-switch\xspace}

\begin{document}

\author{
		Yoni Zohar\inst{1}\and
		Ahmed Irfan\inst{1} \and
		Makai Mann\inst{1} \and
		Andres N\"otzli\inst{1} \and
		Andrew Reynolds\inst{2}\and
		Clark Barrett\inst{1}
}
\institute{
  Stanford University \and
  The University of Iowa
 }

\title{\lazybvtoint at the SMT Competition 2020}

\titlerunning{}
\authorrunning{}

\maketitle

\noindent
\begin{abstract}
\noindent \lazybvtoint is a new prototype SMT-solver, that will participate in
the incremental and non-incremental tracks of the \qfbv logic.
\end{abstract}

\paragraph{Overview.}
\lazybvtoint is a prototype SMT-solver for the theory of fixed-width bit-vectors
and uninterpreted functions. This is the first year it will compete in \smtcomp.
It will participate in the incremental and non-incremental \qfbv tracks.
%
%basic idea
The basic algorithm of the tool relies on a translation from bit-vectors to
non-linear integer arithmetic with uninterpreted functions, followed by a CEGAR
loop~\cite{cegar} that lazily instantiates bit-vector axioms over the
translation. The idea of using integer reasoning for bit-vector solving is not
new (see,
e.g.,~\cite{DBLP:journals/entcs/BozzanoBCFHKPS06,DBLP:conf/fmcad/BackemanRZ18}),
however, it is worth revisiting due to recent improvements in solvers for
non-linear integer arithmetic~\cite{DBLP:conf/vmcai/Jovanovic17,
  DBLP:conf/frocos/ReynoldsTJB17, DBLP:conf/sat/CimattiGIRS18}.
We expect this solver to perform better on benchmarks that involve arithmetic
bit-vector constraints and large bit-widths because the encoding of arithmetic
constraints is straightforward and independent of bit-width, as opposed to the
encoding of bit-wise constraints which is less natural and hindered by larger
bit-widths.
% Makai: minor re-wording, feel free to edit above or switch it back to the text below
% because these seem
% more suitable for the approach of this tool.
%
The tool is open-source and is available at \url{https://github.com/yoni206/lazybv2int}.

\paragraph{Dependencies on Other Tools.}
To parse the input problem, \lazybvtoint employs \msat's
parser through an API~\cite{mathsat5}.
To solve the translated arithmetic problem it uses \cvcfour~\cite{cvc4}.
In some cases, \msat is called on an extension of the original bit-vector problem.
The interface to both external solvers uses
\smtswitch~\cite{smtswitchgithub}.
According to the rules published by the organizers
of \smtcomp 2020, \lazybvtoint is a {\em wrapper tool} (see \cite{rules20})
and not a {\em portfolio solver},
and thus is allowed to compete.
This was confirmed in a private communication with the competition organizers.

%technical details
\paragraph{Technical Details.}
\lazybvtoint works as follows. The input \qfbv formula $\varphi$ is translated
into a \qfufnia formula $\varphi'$. $\varphi'$ is obtained from $\varphi$ by
eliminating bit-vector operators. For arithmetical operators, this is standard.
The bit-wise operators other than bit-wise conjunction, left and right (logical)
shift, and negation are first eliminated in a preprocessing stage using other
bit-vector operators. Bit-wise negation has a standard arithmetical
interpretation which is utilized. Bit-wise conjunction as well as shift
operations are replaced by uninterpreted functions.
The translated formula is solved using a CEGAR-loop that refines the current
formula by adding lemmas. The procedure is complete because the lemma schemes
that are used are complete.
These include some basic properties of the abstracted operations (e.g.,
idempotence of bit-wise \emph{and}), and in the worst case, they include a full
expansion of the operation (using $ite$ operations and summations).
In each step of the loop, an under-approximation check is performed by \msat's
\qfbv solver: the current model $\mu'$ for $\varphi'$ induces assumptions
of the form $x=ToBV(\mu'[x'])$ for variables $x$ of the original formula
and their translations $x'$. 
These assumptions are added in order to solve $\varphi$.
The assumptions do not
include variables which appear in bit-wise \emph{and} and bit-wise \emph{shift} 
operators.
Clearly, $\varphi$ conjoined with the assumptions is an under-approximation of
the original formula $\varphi$. So, if the result of the under-approximation is
is SAT, then $\varphi$ is satisfiable. In case the result of the additional
check is unsatisfiable, then the disjunction of negated assumptions ---only those
assumptions that appear in the unsatisfiable core--- is learned as an additional
refinement lemma.

%conclusion
\paragraph{Conclusion.}
This is a prototype experimental tool that is aimed to serve as a
playground for arithmetic-based techniques for bit-vector solving. Incorporating
such techniques in a full-fledged solver is left for future work, and is planned
for when these techniques are better understood and evaluated using this tool.

%acknowledgments
\paragraph{Acknowledgments.} We would like to thank the \cvcfour and \msat teams for allowing us
to use their tools. In particular, we thank Alberto Griggio
for clarifying the relevant aspects of the \msat license, and to Aina Niemetz
and Mathias Preiner for helpful tips regarding benchmarking and evaluating
bit-vector formulas. We also thank the competition organizers, Haniel Barbosa,
Jochen Hoenicke, and Antti Hyvarinen for clarifying the status of this tool for
the competition and willingly accepting it as a new participant.

  \bibliography{smtcomp-2020}
  \bibliographystyle{plain}

\end{document}